\documentclass[aps,pre,twocolumn,float]{revtex4}
\usepackage{amsmath,bm,epsfig}

\def\Fbox#1{\vskip1ex\hbox to 8.5cm{\hfil\fboxsep0.3cm\fbox{%
  \parbox{8.0cm}{#1}}\hfil}\vskip1ex\noindent}  

\let \= \equiv \let\*\cdot \let\~\widetilde \let\^\widehat \let\-\overline

\begin{document}
\title{Stochastic Processes Crossing from Ballistic to Fractional Diffusion with Memory: Exact Results}
\author{Valery Ilyin$^1$, Itamar Procaccia$^1$ and Anatoly Zagorodny$^2$}
\affiliation{$^1$Department of Chemical Physics, The Weizmann
Institute of Science, Rehovot 76100, Israel\\$^2$ Bogolyubov Institute for Theoretical Physics, 252143 Kiev, Ukraine}
\begin{abstract}
We address the now classical problem of a diffusion process that crosses over from a ballistic behavior at short times
to a fractional diffusion (sub- or super-diffusion) at longer times. Using the standard non-Markovian diffusion equation we demonstrate how
to choose the memory kernel to exactly respect the two different asymptotics of the diffusion process. Having done so
we solve for the probability distribution function (pdf) as a continuous function which evolves inside a ballistically expanding domain.
This general solution agrees for long times with the pdf obtained within the continuous random walk approach but it is much superior to this solution at shorter times where the effect of the ballistic regime is crucial.
\end{abstract}
\maketitle

{\bf Introduction:} Nature offers us a large number of examples of diffusion processes for which an observable $X$
diffuses in time such that its variance grows according to
\begin{eqnarray}
\langle \Delta X^2\rangle (t) \sim D_2 t^2 \quad \mbox{for}~ t \ll t_c \ , \label{var1}\\
\langle \Delta X^2\rangle (t)\sim  D_\alpha t^\alpha\quad  \mbox{for}~ t \gg t_c \ , \label{var2}
\end {eqnarray}
where angular brackets mean an average over repeated experiments and $D_2$ and $D_\alpha$ are coefficients
with the appropriate dimensionality. The short time behavior is known as `ballistic', and is generic for a wide class
of processes. The long time behavior with $\alpha\ne 1$ is generic when the diffusion steps are correlated,
with persistence for $\alpha>1$ and anti-persistence for $\alpha<1$
\cite{MN68}.
These correlations mean that the diffusion
process is not Markovian,  but rather has memory. Thus the probability distribution function (pdf) of the observable
$X$, $f(X,t)$ is expected to satisfy a diffusion equation with memory
\cite{KK74},
\begin{equation}
\frac{\partial f(X,t)}{\partial t} = \int_0^t dt' K(t-t') \nabla^2 f(X,t) \ , \label{pdf}
\end{equation}
with $K(t)$ being the memory kernel and $\nabla^2$ the Laplace operator.

In this Letter we study the class of processes
which satisfy Eqs. (\ref{var1})-(\ref{pdf}).
First of all we find an expression for the kernel $K(t)$ which is unique for a given law of mean-square-displacement. Second we consider the kernel which contains both the ballistic contribution embodied in  Eq. (\ref{var1}) {\em and} the long-time behavior (\ref{var2}). For this case we find an exact equation and a solution for Eq. (\ref{pdf}). Lastly a simple interpolation formula for the kernel is inserted to the exact equation which is then
 solved for the pdf of $X$ without any need for the fractional dynamics approach \cite{MK00}. Some interesting characteristics of the solution
 are described below.

 {\bf Determination of the kernel $K(t)$}: To determine the kernel in Eq. (\ref{pdf}) we use a result obtained in \cite{S02}. Consider the auxiliary equation
 \begin{equation}
 \frac{\partial P(X,t)}{\partial t} =  \nabla^2 P(X,t) \ . \label{pdfP}
 \end{equation}
Define the Laplace transform of the solution of Eq. (\ref{pdfP}) as
\begin{equation}
\tilde P(X,s)\equiv \int_0^t {\rm d}t e^{-st} P(X,t)\ , \label{defLap}
\end{equation}
it was shown in \cite{S02} that the solution of Eq. (\ref{pdf}) with the same initial conditions  can be written as
\begin{equation}
\tilde f(X,s)=\frac{1}{\tilde {K}(s)} \tilde {P}(X,\frac{s}{\tilde{K}(s)}) \ , \label{fxs}
\end{equation}
where here and below the tilde above the symbol means the Laplace transform. The development that we propose here
is to replace in Eq. (\ref{fxs}) the Laplace transform $\tilde K(s)$ with the Laplace transform of the mean-square
displacement. This is done by first realizing (by computing the variance and integrating by parts) that
\begin{equation}
\frac{\partial \langle X^2\rangle(t)}{\partial t}= 2\int_0^t {\rm d}
t^{\prime}K(t-t^\prime)\ , \label{r2K}
\end{equation}
or, equivalently,
\begin{eqnarray}
\tilde K(s) &=& \frac{s^2 \widetilde{\langle {X^2} \rangle}(s)}{2} \ , \label{Ks}\\
&=&\frac{\partial \langle X^2\rangle(t)}{2\partial t} +\frac{s}{2} \widetilde{\frac{\partial \langle X^2\rangle(t)}{\partial t}} - \frac{\partial \langle X^2\rangle(t)}{2\partial t} \ . \nonumber
\end{eqnarray}
The second line was written in order to find the time representation of $K(t)$ which is the inverse Laplace transform:
\begin{eqnarray}
K(t)&=&\frac{1}{2}\Bigg(\delta(t)\frac{\partial}{\partial t}
\langle X^{2}\rangle(t)
 +\frac{\partial^{2}}{\partial t^{2}}\langle X^{2}
\rangle(t)\Bigg) \nonumber \\
&=&\frac{1}{2}\frac{\partial}{\partial t}\Bigg(H(t)
\frac{\partial \langle X^{2}\rangle(t)}{\partial t}\Bigg)\ ,
\label{kern}
\end{eqnarray}
where $H(t)$ is the Heaviside function.  Obviously, using the first line of Eq. (\ref{Ks}) in Eq. (\ref{fxs})
the solution is entirely determined by whatever law is given for the variance, together with initial conditions.

For ordinary diffusion the variance is defined by Eq. (\ref{var2}) with
$\alpha=1$ and $t_c =0$. It follows from Eq. (\ref{kern}) that the kernel is
$K(t)\sim\delta(t)$ and Eq. (\ref{pdf}) is reduced to the Markovian Eq.
(\ref{pdfP}); this process does not possess any memory. More complicated
examples are considered below.


{\bf Example I: fractional differential equations.}
In recent literature the problem of a diffusion process which is consistent
with  Eq. (\ref{var2}) only for all times (i.e. $t_c=0$) is investigated  using the formalism of fractional differential equations (see, e.g., \cite{MK00}).
 In this formalism Eq. (\ref{pdf}) is replaced by the fractional equation
\begin{equation}
\frac{\partial f(X,t)}{\partial t}=D_{\alpha}\hspace{1mm}
_{0}\mathrm{\bf D}^{1-\alpha}_{t} \frac{\partial^2 f(X,t)}{\partial x^2},
\label{frdif}
\end{equation}
where the Rieman-Liouville operator $_{0}\mathrm{\bf D}^{1-\alpha}_{t}$ is
defined by
\begin{equation}
_{0}\mathrm{\bf D}^{1-\alpha}_{t}\phi(x,t)=\frac{1}{\Gamma(\alpha)}
\frac{\partial}{\partial t}\int_0^t{\rm d} t^{\prime}
\frac{\phi(x,t^\prime)}{(t-t^\prime)^{1-\alpha}},
\label{RLop}
\end{equation}
where $\Gamma(\alpha)$ is the gamma function. It is easy to see that this equation follows from Eq. (\ref{pdf}) with the kernel evaluated by Eq. (\ref{kern}) with
the variance (\ref{var2}). We reiterate however that this equation is consistent with Eq. (\ref{var2}) for all times $t\ge 0$. This of course is a problem since this formalism cannot agree with the ballistic short time
behavior which is generic in many systems.

{\bf Example II: ballistic behavior}. For $X$ one-dimensional the solution of Eq. (\ref{pdfP}) with the initial condition $P(X,t=0)=\delta(X)$ is given by
\begin{equation}
\tilde{P}(X,s)=\frac{1}{2\sqrt{ s}}\cdot \exp(-\mid X\mid
 \sqrt{s}) \ .
 \label{onelapl}
\end{equation}
Substituting in Eq. (\ref{fxs}) we find
\begin{equation}
\tilde f(X,s)=\frac{1}{\sqrt{2s^3 \widetilde{\langle X^2\rangle}(s)}}\exp(-\mid X \mid
\sqrt{\frac{2}{s \widetilde{\langle X^2\rangle}(s)}}).
\label{onenloc}
\end{equation}
For systems with the pure ballistic behavior (e.g., dilute gas) the variance can be written as  $\langle X^{2}\rangle_{t}=\langle
u^2\rangle t^2$, where $\langle u^2\rangle$ is the mean square average of
the particle velocities. The Laplace transform of this expression is given by
$\widetilde{\langle X^2 \rangle}(s)=2\langle u^2\rangle/s^3$ and
the Laplace transform of the pdf is defined by
\begin{equation}
\tilde{f}(X,s)=\frac{1}{2\sqrt{\langle u^2\rangle}}~
\exp(-\mid X\mid \frac{s}{\sqrt{\langle u^2\rangle}}) \ . \label{imball}
\end{equation}
The inverse transform reads
\begin{equation}
f(X,t)=\frac{1}{2}\delta(\mid X\mid - \sqrt{\langle u^2\rangle} t) \ . \label{ball}
\end{equation}
This solution corresponds to a deterministic evolution; there is a complete memory of the initial
conditions in the absence of inter-particle interactions, $K(t)=\langle u^2\rangle$).

{\bf General case}: In the general case the mean-square-displacement satisfied some law $\langle X^2\rangle(t)$ which
is supposed to be known at all times, with possible asymptotic behavior as shown in Eqs. (\ref{var1}) and(\ref{var2}).
 To find the appropriate general solution we will split $\tilde f(X,s)$ into two parts, $\tilde f_I(X,s)$ and $\tilde f_{II}(X,s)$, such that the first
part is constructed to agree with the existence of a ballistic regime. Suppose that in that regime, at short time, the mean-square-displacement can be expanded in a Taylor series
\begin{equation}
\langle X^2\rangle (t)=\sum_{i=0}^\infty a_it^{\mu_i-1}\ =a_0t^2+a_1 t^3 +
a_2 t^4\cdots \ , \label{expX2}
\end{equation}
where $\mu_0=3$, $\mu_1=4$ etc. Then the Laplace transform $\widetilde{\langle X^2\rangle}(s)$ can be written for
$s\to \infty$ as \cite{PB64}
\begin{equation}
\widetilde{\langle X^2\rangle}(s) =\sum_{i=0}^\infty a_i \Gamma(\mu_i)\frac{1}{s^{\mu_i}}=2a_0\frac{1}{s^3}+6a_1\frac{1}{s^4}+
24 a_2\frac{1}{s^5}\cdots \ . \label{larges}
\end{equation}
Substituting Eq. (\ref{larges}) up to O$(s^{-4})$ in Eq. (\ref{onenloc}) yields
\begin{equation}
\tilde{f}_{I}(X,s)_{s\to\infty}=\frac{1}{2\sqrt{a_0}}
exp(-\frac{\mid X\mid}{\sqrt{a_{0}}}(s-\frac{3a_1}{2a_0})) \ . \label{aspdf}
\end{equation}
The inverse Laplace transform of this result reads
\begin{equation}
f_{I}(X,t)=\frac{1}{2}exp(\frac{3a_1}{2a_0}t)\delta(\mid X \mid-\sqrt{a_{0}}t) \ . \label{f1}
\end{equation}
Not surprisingly, this partial solution corresponds to a deterministic propagation. Note that in order to avoid
exponential divergence in time we must have $a_{1}<0$ in the expansion (\ref{expX2}).

Having found $\tilde f_{I}(X,s)$ we can now write $\tilde f_{II}(X,s)$ simply as
\begin{equation}
\tilde f_{II}(X,s) = \tilde f(X,s) - \tilde f_{I}(X,s) \ .
\end{equation}
Calculating this difference explicitly we find
\begin{widetext}
\begin{eqnarray}
\tilde{f}_{II}(X,s)&=&\frac{1}{2}\Bigg(\sqrt{\frac{2}{s^3 \widetilde{\langle X^2\rangle}(s))}}
\exp(-\mid X \mid(\sqrt{\frac{2}{s\widetilde{\langle X^2\rangle}(s)}}-\frac{s}{\sqrt{a_{0}}}))
-\frac{1}{\sqrt{a_0}}
\exp(\frac{3a_{1}}{2a_{0}^{3/2}}\mid X\mid)\Bigg) \exp(-\frac{\mid X\mid}
{\sqrt{a_{0}}}s)
\nonumber \\
&\equiv&\tilde{F}(X,s)\exp(-\frac{\mid X\mid}{\sqrt{a_{0}}}s).
\label{f2}
\end{eqnarray}
\end{widetext}
The inverse Laplace transform of Eq. (\ref{f2}) is given  by
\begin{equation}
f_{II}(X,t)=F\Big(X,t-{\mid X\mid}/{\sqrt{a_0}}\Big)H(\sqrt{a_{0}}t-\mid X\mid).
\label{orf2}
\end{equation}

The importance of this result is that the explicit Heaviside function is taking upon itself
the discontinuity in the solution $f_{II}(X,t)$. The exact value of this
function at the point $\mid X\mid=\sqrt{a_{0}}t$ can be calculated using the
initial value theorem and is given by
\begin{eqnarray}
&&f_{II}(\mid X\mid=\sqrt{a_0}t,t)=\label{invalue} \\&&
-\Bigg(\frac{3}{4}\frac{a_1}{a_{0}^{3/2}}+\frac{1}{2\sqrt{a_0}}(\frac{27}{8}
[\frac{a_1}{a_0}]^2-6\frac{a_2}{a_0})t\Bigg) 
\times\exp(-\frac{3}{2}\frac{a_1}{a_0}t) \ . \nonumber
\end{eqnarray}

Summing together the results (\ref{f1}) and (\ref{orf2}) in the time domain we get a general
solution of the non-Markovian problem with a short-time ballistic behavior, in the form
\begin{eqnarray}
f(X,t)&=&\frac{1}{2}exp(\frac{3a_1}{2a_0}t)\delta(\mid X\mid-\sqrt{a_{0}}t)
\nonumber \\
&+&F(X,t-\frac{\mid X\mid}{\sqrt{a_0}})H(\sqrt{a_{0}}t-\mid X\mid).
\label{fullsol}
\end{eqnarray}
This is the main result of the present Letter.
The diffusion repartition of the probability distribution function occurs
inside the spatial diffusion domain which increases in a deterministic way.
\begin{figure}[!h]
\centering
\includegraphics[width=0.40\textwidth]{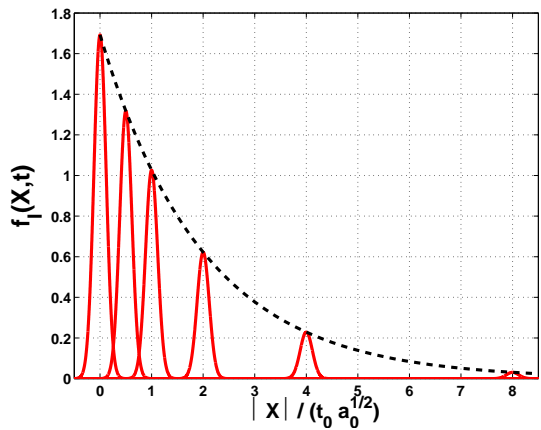}
\caption{  The time evolution of the function $f_{I}(X,t)$ defined by Eq.
(\ref{f1}) for time intervals  $t/t_{0}=$0.5, 1, 2, 4, 8 (the time scale
$t_0=a_0/(3 a_1)$). The $\delta$-function is graphically represented by narrow
Gaussians.}
\label{fig1}
\end{figure}
The first term in Eq. (\ref{fullsol}) corresponds to the propagating $\delta$-function
which is inherited from the initial conditions, and it lives at the edge of the ballistically expanding domain. Schematically the time evolution
of this term is shown in Fig.\ref{fig1}, where the $\delta$-function is graphically represented as a narrow Gaussian.
The dashed line represents the exponential decay of the integral over the $\delta$-function. The function $F(X,t-\frac{\mid X\mid}{\sqrt{a_0}})$ in the time domain is a continuous
function  and can be evaluated numerically, for example using the direct
integration method \cite{duf93}. Below we demonstrate this calculation
with explicit examples.

{\bf Interpolation for all times}: To interpolate Eqs. (\ref{var1}) and (\ref{var2}) we propose the form
\begin{equation}
\langle\Delta X^{2}\rangle_t=2D_{\alpha}t_{0}^{\alpha}\frac{(t/ t_{0})^{2}}
{(1+(t / t_{0}))^{2-\alpha}},
\label{gdisp}
\end{equation}
where $0\leq \alpha\leq 2$. Here $t_{0}$ is the crossover characteristic time,
at $t\ll t_{0}$ the law (\ref{gdisp}) describes the ballistic regime and at
$t\gg t_{0}$ the fractional diffusion.

\begin{figure}[!h]
\centering
\includegraphics[width=0.35\textwidth]{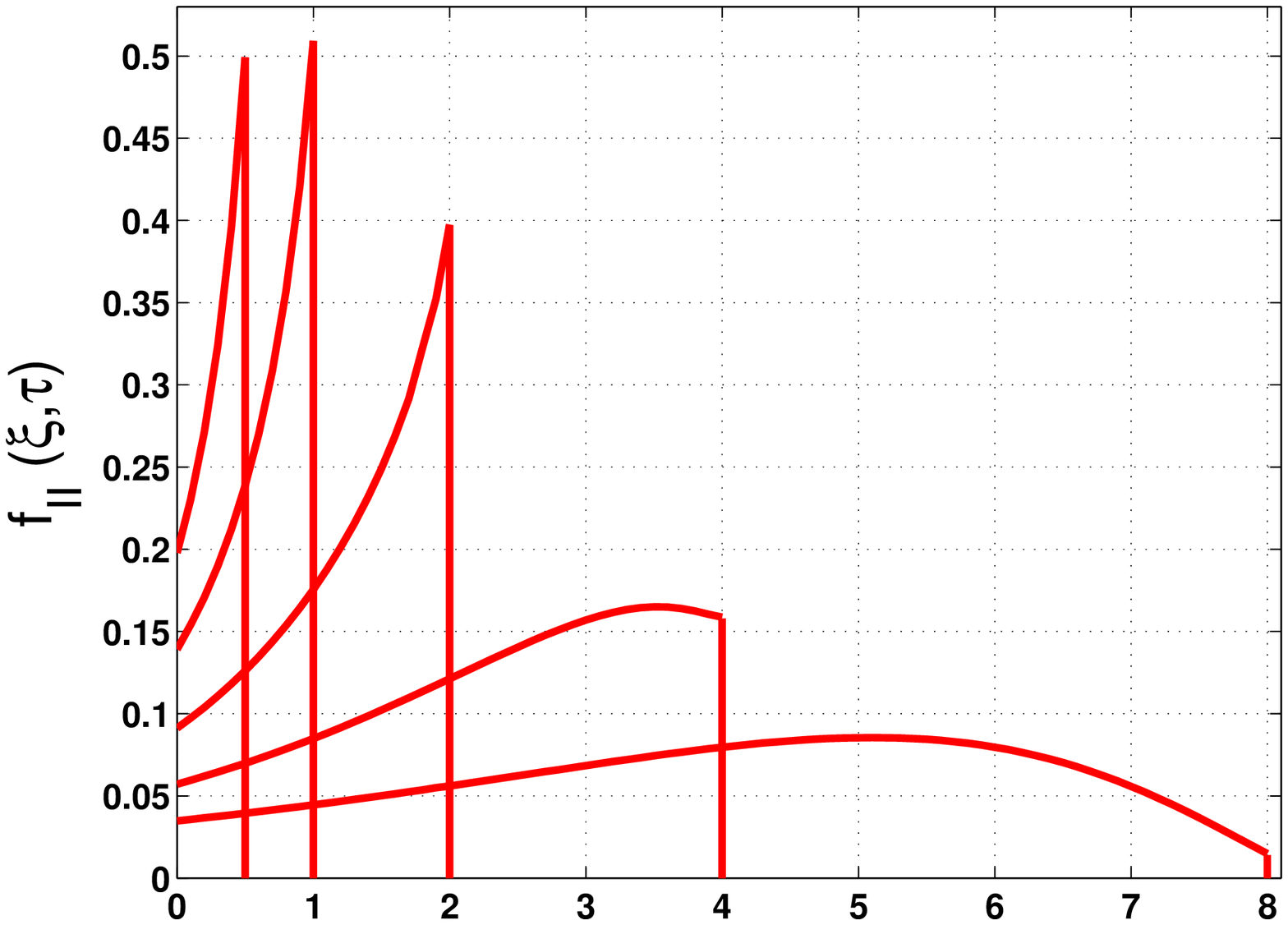}
\includegraphics[width=0.35\textwidth]{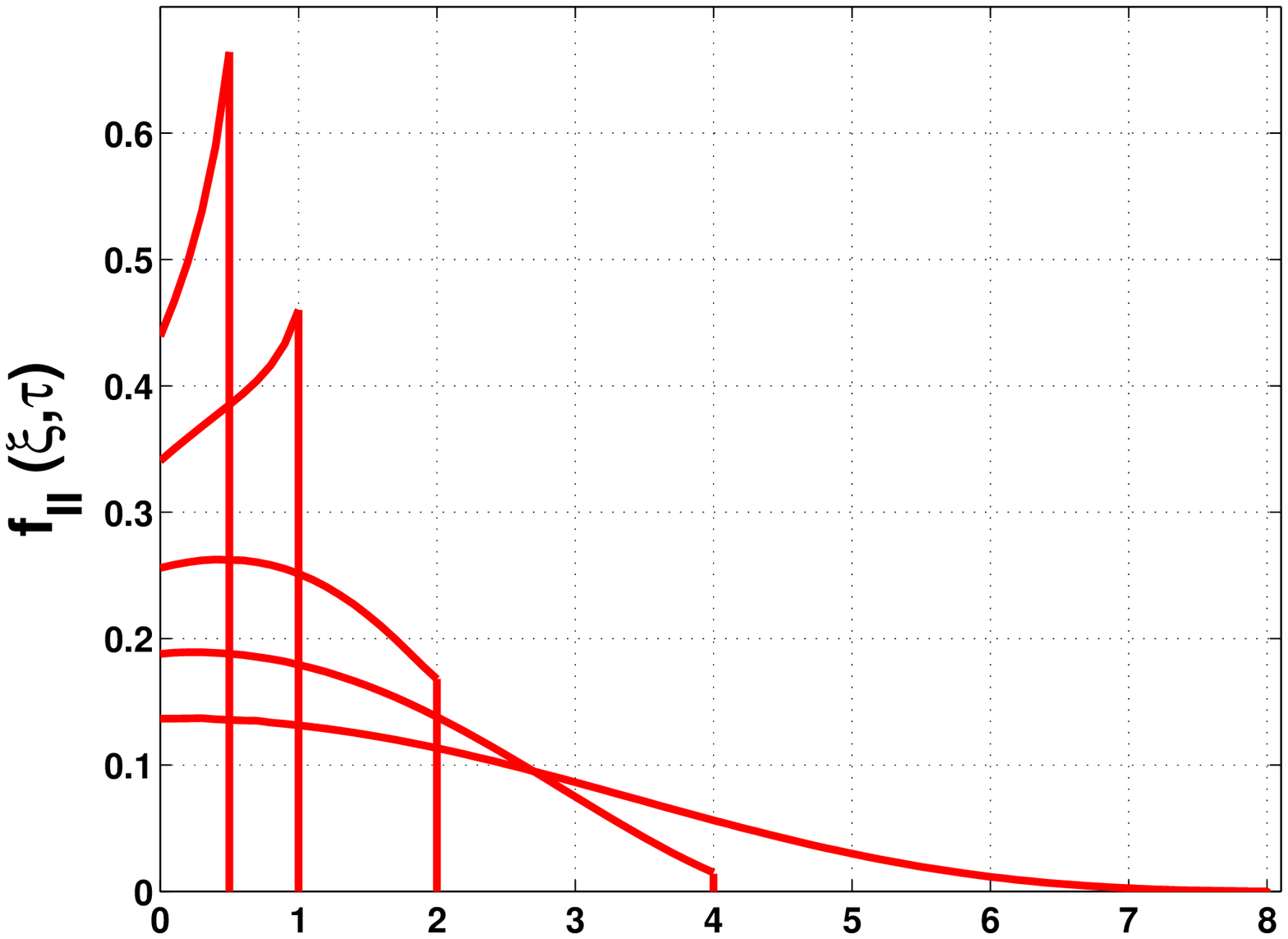}
\includegraphics[width=0.35\textwidth]{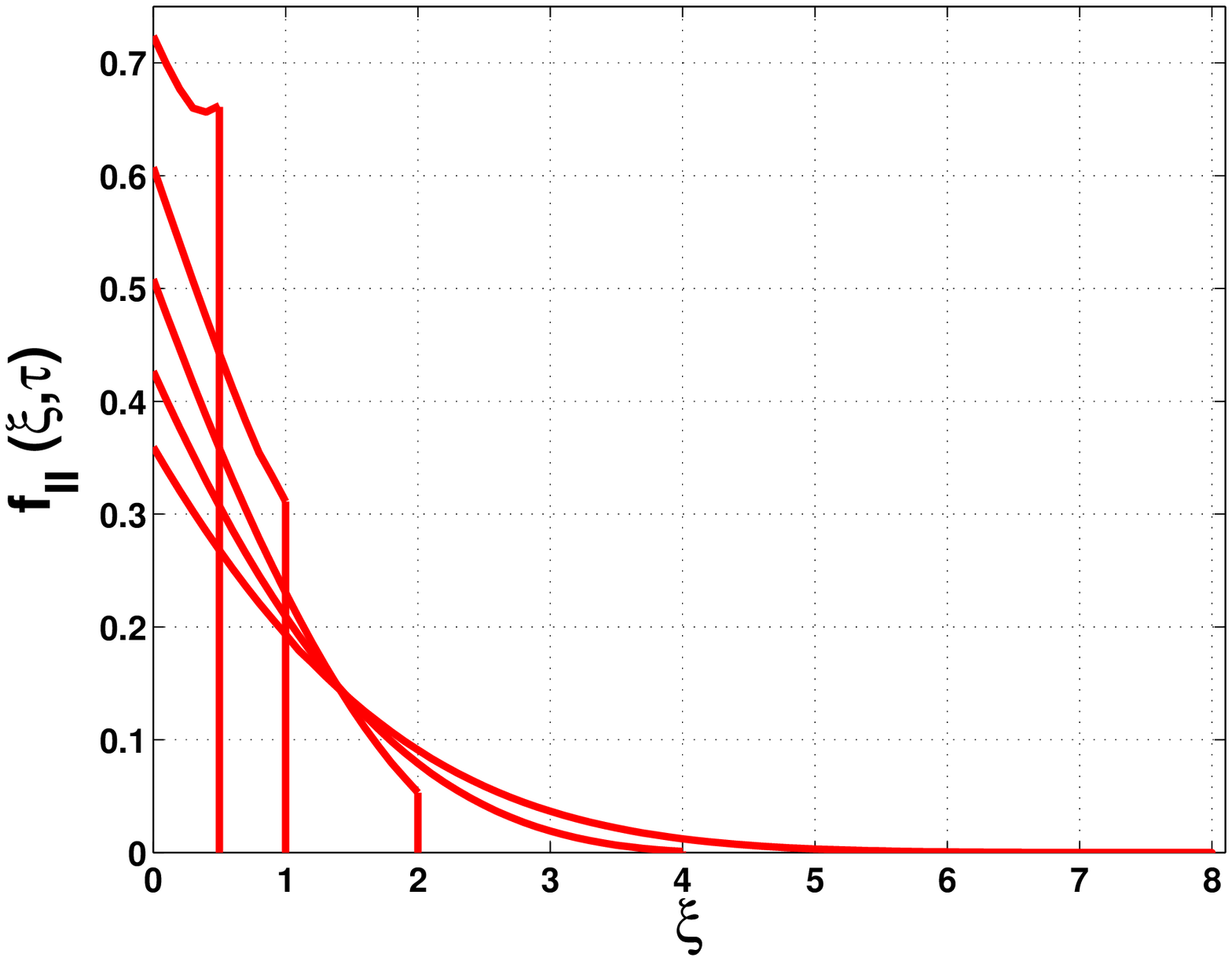}
\caption{The continuous part of the pdf
(\ref{orf2}) for different values of the parameter $\alpha$. Superdiffusion
($\alpha=3/2$, upper panel), regular diffusion ($\alpha=1$, middle panel) and
subdiffusion ($\alpha=1/2$, lower panel).  Time intervals from the top to the
bottom $\tau=$0.5, 1, 2, 4, 8. The reader should note that the full solution of the problem
is the sum of the two solutions shown in this and the previous figure.}
\label{fig2}
\end{figure}

Introduce now dimensionless variables $\langle\xi^2\rangle _\tau =
\langle\Delta X^{2}\rangle _{t}/(2 D_{\alpha}t_{0}^{\alpha})$ and $\tau=t/t_0$. With these
variables the last equation reads
\begin{equation}
\langle\xi^2\rangle _\tau =\frac{\tau^2}{(1+\tau)^{2-\alpha}}.
\label{gdispd}
\end{equation}
The Taylor expansion of (\ref{gdisp}) is given by
\begin{equation}
\langle\xi^2\rangle _\tau =\tau^2-(2-\alpha)\tau^3+\frac{1}{2} (3-\alpha)(2-\alpha) \tau^4+\ldots .
\label{gtayl}
\end{equation}
Substitution these expansion coefficients into Eq.~(\ref{f1}) yields the first term in
the expression for the probability distribution function (\ref{fullsol})
\begin{equation}
f_{I}(x,t)=\frac{1}{2}exp(-\frac{3(2-\alpha)}{2}\tau)
\delta(\mid \xi \mid-\tau).
\label{gf1}
\end{equation}

The Laplace transform of Eq. (\ref{gdispd}) is
\begin{equation}
\widetilde{\langle X^2\rangle}(s)=(\frac{\alpha}{s}-1)\frac{1}{s}+
\bigg( (\alpha-1)(\frac{\alpha}{s}-2)+s\bigg)\frac{e^s}{s^\alpha}
\Gamma(\alpha-1,s),
\label{img}
\end{equation}
where $\Gamma(a,s)$ is the incomplete gamma function.
Note that the case $\alpha=2$ is special, since it annuls the exponent in Eq. (\ref{gf1}), leaving
  as a solution a ballistically propagating $\delta$-function.  For all other values
  of $\alpha<2$ the inverse
Laplace transform of the function $\widetilde{F}(x,s)$ which defines the
diffusion process inside the expanded spatial domain should be evaluated, in general,
numerically.

Results of the calculations following the method of Ref. \cite{duf93} for the smooth part of the probability distribution function $f_{II}(x,t)$ for different values of the parameter $\alpha$ are
shown in Fig. \ref{fig2}. The reader should appreciate the tremendous role of memory, or the non-Markovian
nature of the process under study. For example regular diffusion with $\alpha=1$ results in a Gaussian pdf
that is peacefully expanding and flattening as time increases. Here, in the mid panel of Fig. \ref{fig2} we see
that the ballistic part which is represented by the advancing and reducing $\delta$-function sends backwards the probability that it loses due to the exponential decay seen in Fig. \ref{fig1}. This `back-diffusion' leads initially to a qualitatively different looking pdf, with a maximum at the edge of the ballistically expanding domain. At later times
the pdf begins to resemble more regular diffusion. The effect strongly depends on $\alpha$ simply due to the
appearance of $\alpha$ in the exponent in Eq. (\ref{gf1}).

For long times the solutions shown in Fig. \ref{fig2} agree with
the Markovian pdf obtained in the frame of a continuous-time random walk
\cite{BHW87}. For the special case $\alpha=0$ the limiting behavior of the
general solution from Eq. (\ref{fullsol}) can be evaluated with the help of
the final value theorem:
\begin{equation}
f(X)=\frac{1}{\sqrt{2}} \exp(-\sqrt{2}\mid X\mid),
\label{azer}
\end{equation}
this analytical result coincides with the pdf from \cite{BHW87} at the same
conditions.

In summary, we have shown how to deal with diffusion processes that cross-over from
a ballistic to a fractional behavior for short and long times respectively, within
the time non-local approach. The general solution (\ref{fullsol}) demonstrates
the effect of the temporal memory in the form of a partition of the probability distribution function
inside a spatial domain which increases in a deterministic way. The approach provides a solution that
is valid at all times, and in particular is free from  the instantaneous action puzzle.

\end{document}